\documentclass[aps,prb,twocolumn,showpacs,superscriptaddress]{revtex4}
\usepackage{graphicx}
\usepackage{amsfonts,amssymb,amsmath}
\usepackage{bm}
 
\begin{document}
 
\title{Magnetism in systems with various dimensionality: A comparison
between Fe and Co}

\date{\today}
 
\author{Claude Ederer}
\affiliation{Max-Planck Institut f\"ur Metallforschung, Heisenbergstr.~3,
  D-70569 Stuttgart, Germany}
\author{Matej Komelj}
\affiliation{Jo\v zef Stefan Institute, Jamova 39, SI-1000 Ljubljana,
  Slovenia}
\author{Manfred F\"ahnle}
\email{faehnle@mf.mpg.de}
\homepage{http://physix.mpi-stuttgart.mpg.de/schuetz/elth/electronth.html}
\affiliation{Max-Planck Institut f\"ur Metallforschung, Heisenbergstr.~3,
  D-70569 Stuttgart, Germany}

\begin{abstract}
A systematic {\it ab initio} study is performed for the spin and
orbital moments and for the validity of the sum rules for x-ray
magnetic circular dichroism for Fe systems with various dimensionality
(bulk, Pt-supported monolayers and monatomic wires, free-standing
monolayers and monatomic wires). Qualitatively, the results are similar
to those for the respective Co systems, with the main difference that
for the monatomic Fe wires the $\langle T_z \rangle$ term in the spin
sum rule is much larger than for the Co wires. The spin and orbital
moments induced in the Pt substrate are also discussed.
\end{abstract}

 
\maketitle
 
One of the outstanding features of magnetism in low-dimensional spin
systems is the importance of the spin-orbit-coupling effects.  Whereas
the orbital moments in bulk materials of high symmetry are strongly
reduced due to orbital quenching, they may attain large values in
low-dimensional systems. For instance, in monatomic Co wires at the
steps of a vicinal Pt surface an orbital moment of about
0.68$\mu_\text{B}$ per Co atom was found \cite{Gambardella:2002},
which is a factor of about 5 larger than in bulk hcp Co and which
represents the largest orbital moment ever reported for a 3$d$
itinerant electron system. For the magnetic-anisotropy energy of this Co wire
a value was deduced \cite{Gambardella:2002} which is about 50 times
larger than the one of hcp Co (which is already large). The large
magnetic anisotropy is also relevant for the nature of magnetic
excitations. It will generate an excitation gap for the linear
excitations, i.e., the spin waves, and for one-dimensional systems it
will lead to nonlinear excitation modes which possibly have the
character of solitons \cite{Mikeska:1991}.
\par 

A suitable experimental method to resolve spin and orbital moments is
the technique of x-ray magnetic circular dichroism (XMCD). This
technique \cite{Schuetz:1987} is based on the fact that for magnetic
systems the absorption coefficient $\mu(\epsilon)$ as function of the
x-ray energy $\epsilon$ is different for x-rays with left-circular
polarization, $\mu_+(\epsilon)$, and right-circular polarization,
$\mu_-(\epsilon)$. For example, in a $3d$ transition metal system one
has to measure these absorption coefficients at the $L_2$ and $L_3$
edge corresponding to the $3d$ transition metal atoms in the
investigated system, and one then can obtain the orbital moment $m_l =
- \mu_\text{B} \langle l_z \rangle$ and the spin moment $m_s = -
\mu_\text{B} \langle \sigma_z \rangle$ for a transition metal atom via
the XMCD sum rules \cite{Thole:1992,Carra:1993} according to:

\begin{equation}
\label{orbsumrule} 
\langle l_{z} \rangle =  
\frac{2I_{\text{m}}N_{\text{h}}}{I_{\text{t}}} \quad ,
\end{equation}
\begin{equation}
\label{spinsumrule} 
\langle \sigma_{z}\rangle = 
\frac{3I_{\text{s}}N_{\text{h}}}{I_{\text{t}}} 
- 7\langle T_{z} \rangle \quad ,
\end{equation}
\begin{equation}
I_{\text{m}} = \int\limits_{E_{\text{F}}}^{E_{\text{c}}} \left[
(\mu_{\text{c}})_{L_3} + (\mu_{\text{c}})_{L_2} \right] d\epsilon
\quad ,
\end{equation}
\begin{equation}
I_{\text{s}} = \int\limits_{E_{\text{F}}}^{E_{\text{c}}} \left[
(\mu_{\text{c}})_{L_3} - 2(\mu_{\text{c}})_{L_2} \right] d\epsilon
\quad ,
\end{equation}
\begin{equation}
I_{\text{t}} = \int\limits_{E_{\text{F}}}^{E_{\text{c}}} \left[
(\mu_{\text{t}})_{L_3} + (\mu_{\text{t}})_{L_2} \right] d\epsilon
\quad ,
\end{equation}
with the XMCD signal $\mu_\text{c} = \mu^+ - \mu^-$ and with
$\mu_\text{t} = \mu^+ + \mu^- + \mu^0$. Here we have assumed that the
$z$-axis is parallel to the propagation direction of the x-rays. Then
$\mu_0(\epsilon)$ is the absorption coefficient for linear
polarization along the $z$-axis. $N_\text{h}$ is the number of holes
in the $d$ part of the valence band and $\langle T_z \rangle$ is the
expectation value of the magnetic dipolar operator
\begin{equation}
\hat{T}_{z} = \frac{1}{2} [ \bm{\sigma} -
3\hat{\mathbf{r}}(\hat{\mathbf{r}}\cdot\bm{\sigma}) ]_{z} \quad ,
\end{equation}
where $\bm{\sigma}$ denotes the vector of the Pauli matrices. The
quantities $E_{\text{F}}$ and $E_{\text{c}}$ denote the Fermi energy
and a cutoff energy. For details of such calculations see
Ref.~\onlinecite{Ederer:2002}.  
\par 
For a practical application of these sum rules there are several
problems, especially for low-dimensional systems, and we just want to
mention two of them. First, the $\langle T_z \rangle$ term in general
is not known from the experiment, and it is therefore often
neglected. However, it is a measure of the anisotropy of the spin
density in the material, and it is therefore expected that it becomes
more and more important when going to systems with more and more
reduced dimensionality and hence more and more reduced
symmetry. St\"ohr and K\"onig \cite{Stoehr/Koenig:1995} introduced an
angle-dependent XMCD technique which allows to eliminate the $\langle
T_z \rangle$ term from the analysis with the spin sum rule. The basic
idea is to measure $\mu_\text{c}(\epsilon)$ for an orientation of the
magnetization in $x$, $y$, and $z$ direction. For systems with weak
spin-orbit coupling we have $\langle T_x \rangle + \langle T_y \rangle
+ \langle T_z \rangle \approx 0$, and if this precondition is
fulfilled then the contribution of the magnetic dipolar term drops out
when we average the spin sum rule over the three directions of the
magnetization. However, for low-dimensional systems the effects of
spin-orbit coupling are larger, and it might be that then the
precondition for the application of the angle-dependent XMCD is no
longer fulfilled. We have shown this explicitely for the extreme
situation of monatomic Fe, Co, and Ni wires \cite{Ederer:2003}. A
second problem is that a couple of assumptions and approximations had
to be made to derive the XMCD sum rules (see, e.g.,
Ref.~\onlinecite{Ederer:2002} and references therein) the validity of
which is not guaranteed in all systems. The validity of the XMCD sum
rules has been confirmed, e.g., for bulk Fe and Co
\cite{Chen_et_al:1995}, but for atoms at free surfaces \cite{Wu:1994}
and at interfaces \cite{Ederer:2002}, i.e., for sites with low
symmetry, the application of the sum rules may be critical. It is
therefore important to figure out the validity of the sum rules for
low-dimensional systems.
\par 
In Ref.~\onlinecite{Komelj:2002} we performed a systematic {\it ab
initio} study for the spin and orbital moments and for the validity of
the XMCD sum rules in Co systems with various dimensionality, i.e.,
for hcp Co, a Co monolayer on Pt (111) and a free standing Co (111)
monolayer, a Co monatomic wire at the steps of a Pt (111) surface as
well as for a free-standing monatomic Co wire. In the present Brief
Report we discuss the respective data for the case of
Fe. Corresponding XMCD measurements are on the way for bulk Fe, an Fe
monolayer on Pt and a monatomic Fe wire on Pt \cite{Enders:private}.
\par
\begin{figure}
\includegraphics[width=0.4\textwidth]{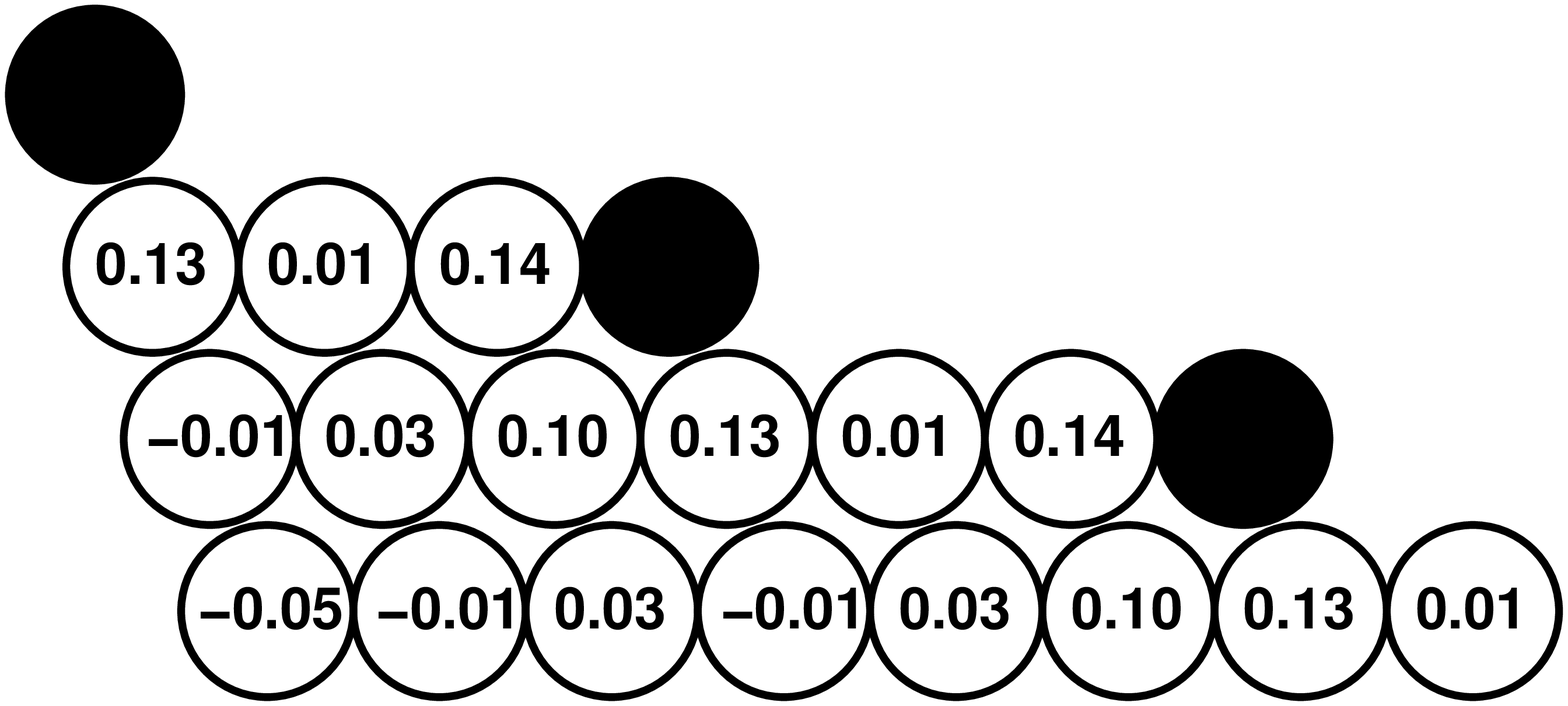}
\includegraphics[width=0.4\textwidth]{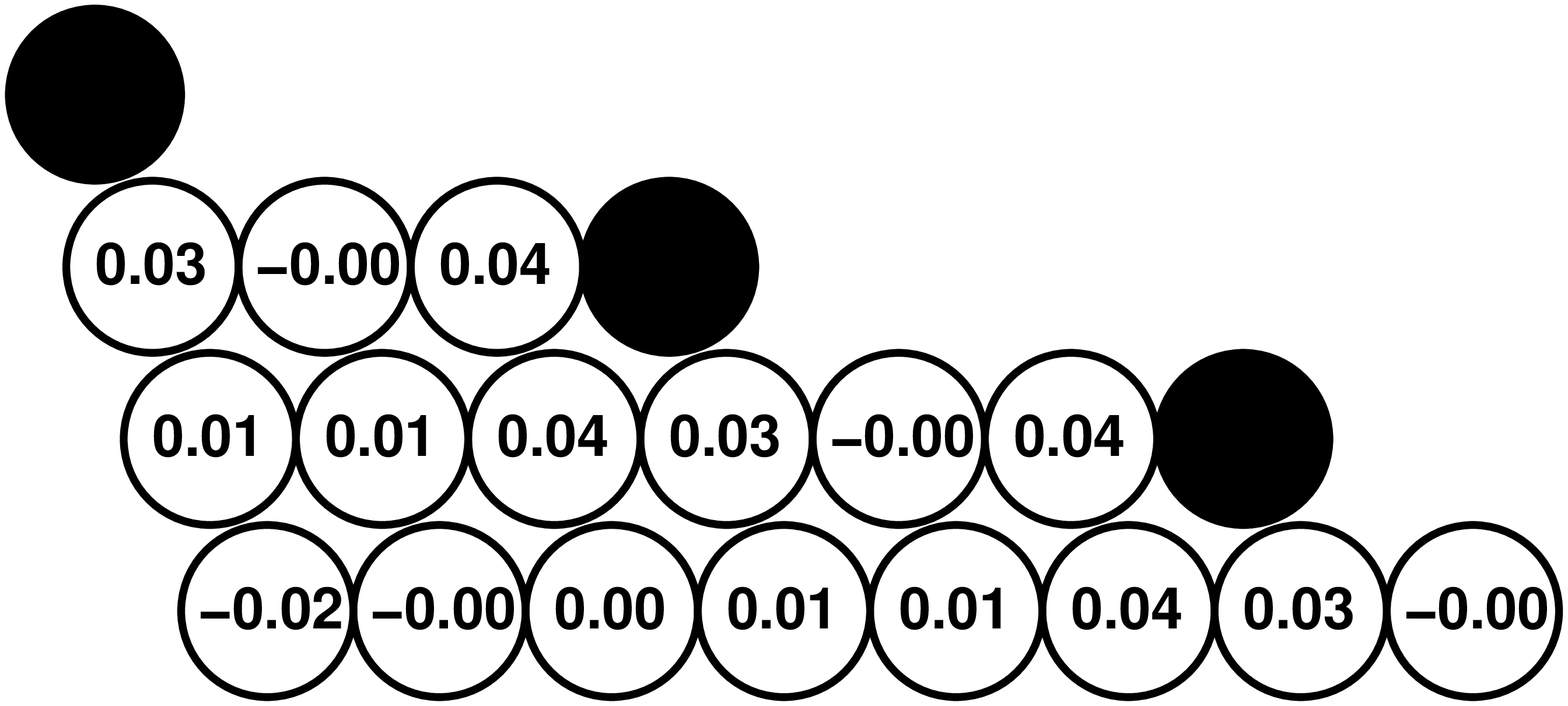}
\caption{The spin (upper graph) and orbital (lower graph) moments
induced on the Pt atoms (open circles) by the monatomic Fe chains
(full circles) at the steps of a vicinal Pt surface.}
\label{Fe:induced}
\end{figure}

 
The validity of the sum rules was tested by calculating the orbital
and spin moments on the one hand directly by the {\it ab initio}
density functional electron theory, on the other hand the absorption
spectra were determined by the same {\it ab initio} theory and then a
second set of orbital and spin moments was obtained via the sum
rules. Deviations between the two sets of data show that the sum rules
are violated. We performed {\it ab-initio} calculations by using the
local-spin-density approximation \cite{Perdew:1992} (LSDA) in a
combination with spin-orbit coupling, without and with an
orbital-polarization (OP) term \cite{Eriksson:1990}. The orbital
polarization term corrects at least in part explicitely for those
electronic correlations which are responsible for the orbital
polarization effects (in free atoms described by Hund's second rule)
and which are not appropriately described by the LSDA. We used the
tight-binding linear-muffin-tin-orbital (LMTO) method in the atomic
sphere approximation \cite{Andersen:1984}, and the WIEN97 code
\cite{Blaha:1990} which adopts the full-potential
linearized-augmented-plane-wave (FLAPW) method \cite{Wimmer:1981} in
which the spin-orbit coupling and the tools for the calculation of the
XMCD spectra \cite{Kunes:2001-1,Kunes:2001-2} and the orbital
polarization term \cite{Rodriguez:2001} have been implemented. A
supercell geometry \cite{Komelj:2002} was used for all calculations
with perpendicular magnetization for the case of the monolayers and
wires. For the monolayer the supercell consists of two Pt and one Fe
(111) layers in the fcc stacking and a vacuum sheet on top of the Fe
layer corresponding to two empty layers. The vicinal Pt surface with
the Fe wires at the steps is modelled by the supercell shown in
Fig.~\ref{Fe:induced}.  To focus on the pure effect of the
dimensionality we fixed the nearest-neighbor distances of the atoms
for all considered systems to the one of fcc Pt (2.77\AA), as it was
done in the LSDA study of finite Co chains on a plane Pt (111) surface
by Lazarovits {\it et al.}  \cite{Lazarovits:2003}. (In contrast,
Spi{\v s}ak and Hafner \cite{Spisak/Hafner:2002} have taken into
account the structural relaxation effects in their LSDA study of
monatomic Fe wires on vicinal Cu surfaces.)
\par
\begin{figure}
\includegraphics[width=0.4\textwidth]{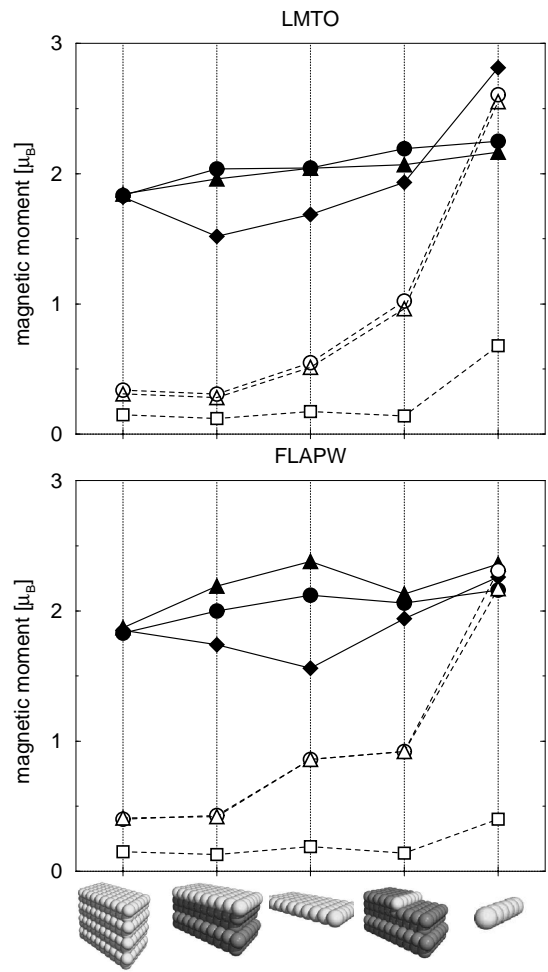}
\caption{The results for the spin moments (upper three curves) and for
the orbital moments (lower three curves) for Co systems of various
dimensionality. For the meaning of the symbols see text. The upper
graph shows the LMTO results, the lower graph shows the FLAPW
results.}
\label{Co:results}
\end{figure}
Fig.~\ref{Co:results} summarizes the results for the Co systems. For
the spin moments nearly the same results were obtained when taking
into account the OP term or when neglecting it. We therefore only show
the data obtained from calculations with the OP term. Six types of
calculations were performed:
\begin{enumerate}
\item{A direct calculation of the spin moments (filled circles connected
by full lines).}
\item{A calculation of the spin moments based on the spin sum rule,
thereby including the $\langle T_z \rangle$ term (filled triangles
connected by full lines).}
\item{A calculation of the spin moments based on the spin sum rule,
thereby neglecting the $\langle T_z \rangle$ term (filled diamonds
connected by full lines).}
\item{A direct calculation of the orbital moments in LSDA (open
squares connected by dashed lines).}
\item{A direct calculation of the orbital moments in LSDA+OP (open
circles connected by dashed lines).}
\item{A calculation of the orbital moments in LSDA+OP based on the
orbital sum rule (open triangles connected by dashed lines).}
\end{enumerate}
\par
\begin{figure}
\includegraphics[width=0.4\textwidth]{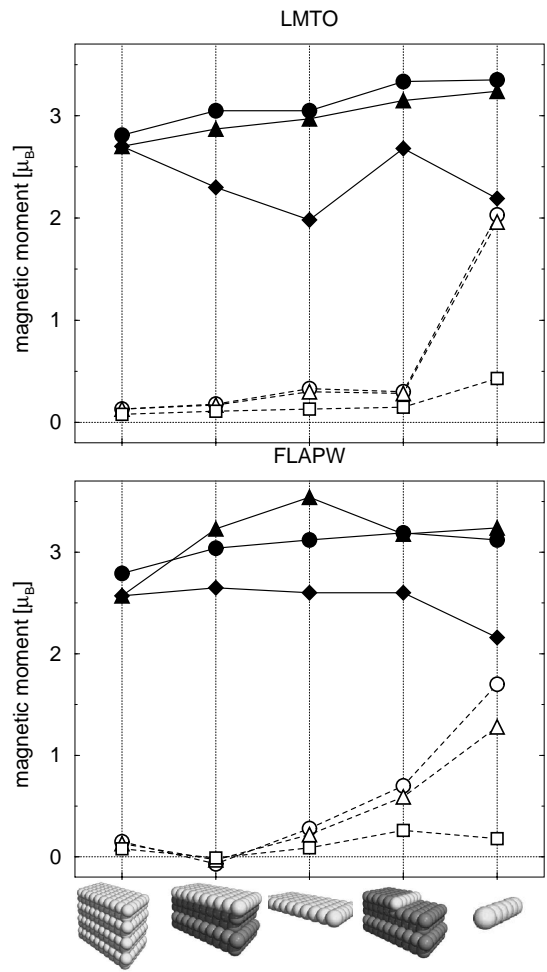}
\caption{The results for the spin moments (upper three curves) and for
the orbital moments (lower three curves) for Fe systems of various
dimensionality. For the meaning of the symbols see text.The upper
graph shows the LMTO results, the lower graph shows the FLAPW
results.}
\label{Fe:results}
\end{figure}

The comparison of the results for the LMTO and the FLAPW calculations
given in Figs.~\ref{Co:results},\ref{Fe:results} shows that the
general trends are essentially the same for both types of
calculations. Quantitatively, there are some differences which result
mainly from the fact that the LMTO method adopts a spherical
approximation for the effective potential in each atomic sphere (ASA
potential) whereas in the FLAPW calculation the full asphericity of
the effective potential is taken into account. In the LMTO-ASA method
the $\langle T_z \rangle$ term is determined from the nonspherical
charge and spin density obtained after the last iteration
step. Because this charge and spin density is calculated from an ASA
potential, the influence of the nonspherical parts of the effective
potential within the spheres on the asphericity of the charge and spin
density is neglected. Therefore, for a quantitative discussion the
FLAPW results are more reliable.
\par
The most important results for the Co systems are:
\begin{enumerate}
\item[a.]{The spin moments increase only slightly with decreasing
dimensionality.}
\item[b.]{The orbital moments increase strongly with decreasing
dimensionality when we take into account the OP term.}
\item[c.]{The $\langle T_z \rangle$ term appearing in the spin sum rule is
only relevant for the monolayers, otherwise it is
rather small (see the more realistic FLAPW data), even for the
monatomic wires. It has been outlined 
in Ref.~\onlinecite{Komelj:2002} that this results from the special
band filling for the case of Co wires.}
\item[d.]{Both the spin sum rule and especially the orbital sum rule
are rather well fulfilled for all Co systems, irrespective of the
dimensionality.}
\end{enumerate}
\par
Fig.~\ref{Fe:results} shows the corresponding results for the case of
Fe. Again we want to focus on the pure influence of the dimensionality
and we therefore fixed the nearest-neighbor distances of the atoms for
all the considered systems to the one of fcc Pt. For bulk Fe we
considered the hypothetical ferromagnetic fcc phase. The four most
important results are:
\begin{enumerate}
\item[a.]{The spin moments increase only slightly with decreasing
dimensionality, as in the case of Co.}
\item[b.]{The orbital moments increase strongly with decreasing
dimensionality when we take into account the OP term, as in the case
of Co.}
\item[c.]{The contribution of the $\langle T_z \rangle$ term to the spin
sum rule is considerable for the monatomic layers and also for the
monatomic wires. This will represent a problem for the determination
of the spin moment from experimental XMCD spectra for Fe monolayers
and Fe monatomic wires: The $\langle T_z \rangle$ term cannot be
neglected in the spin sum rule, and --- at least for the monatomic
wires --- it cannot be determined safely from angle-dependent XMCD
measurements because the above discussed precondition for the
application of this technique is not fulfilled
\cite{Ederer:2003}.}
\item[d.]{There is a tendency that the spin and orbital sum rules are a bit
less well fulfilled than for the Co systems, but the violations are
still moderate.}
\end{enumerate}
\par
For an interpretation of the experiments it is also interesting to
know the spin and orbital moments induced by the Co or Fe atoms in the
Pt substrate, because the Pt atoms contribute to any magnetic
property of the system, for instance, to the total magnetization and
especially to the magneto-optical properties in the visible regime of
light. Experimentally, they can be separated from the spin and orbital
moments of the Co and Fe atoms by the XMCD technique
\cite{Schuetz:1987}.
\begin{table}
\begin{tabular}{|c||c|c|c|c||c|c|c|c|}
\hline
 & \multicolumn{4}{c||}{Co} & \multicolumn{4}{c|}{Fe} \\
\hline
 & \multicolumn{2}{c|}{LMTO} & \multicolumn{2}{c||}{FLAPW} &
 \multicolumn{2}{c|}{LMTO} & \multicolumn{2}{c|}{FLAPW}  \\
\hline
 & spin & orbital & spin & orbital & spin & orbital & spin & orbital
 \\
\hline\hline
1. Pt layer & 0.28 & 0.08 & 0.28 & 0.10 & 0.23 & 0.07 & 0.24 & 0.04 \\
\hline
2. Pt layer & 0.13 & 0.03 & 0.15 & 0.05 & 0.06 & 0.01 & 0.11 & 0.02 \\
\hline
\end{tabular}
\caption{Results for the induced spin and orbital moments for the Pt
supported monolayers obtained by the LSDA+OP calculations.}
\label{Tabelle}
\end{table}
Table \ref{Tabelle} shows the results obtained from the LSDA
calculations for the Pt supported monolayers including the OP term
(very similar results are obtained for the pure LSDA calculation). For
the Fe monolayer the induced Pt moments are a bit smaller than for the
Co monolayer, especially for the second Pt layer. 
It should be recalled that no relaxation effects are taken into
account and that because of the supercell geometry there is a vacuum
sheet below the second Pt layer. Therefore, the quantitative results
should not be taken too literally when comparing with experimental
data, but the calculations certainly yield the correct order of
magnitude for the polarization effect. A similar discussion holds also
for the case of the wires on Pt, for which the results for the
monatomic Fe wire obtained by the LMTO method are given in
Fig.~\ref{Fe:induced}. It is interesting to note that for this
geometry we get nearly exactly the same results as for the case of a
monatomic Co wire.  
\par 
Altogether, the {\it ab-initio} electron theory does not predict a
tremendous difference in the physics of Co and Fe monatomic wires
although an eventual application of the XMCD spectroscopy on the Fe
wires may be less reliable, particularly due to the discussed
importance of the magnetic dipolar term.

\bibliography{feco.bib}

\end{document}